\documentclass[aps,prb,groupedaddress,superscriptaddress,twocolumn]{revtex4}

\usepackage{graphicx, amsmath, subfigure}
\usepackage{amsfonts}
\usepackage{hyperref}
\usepackage{braket}
\usepackage{color}
\usepackage{float}
\usepackage{chngcntr}
\usepackage{mathrsfs}
\newcommand\crbeup[1]{b_{#1, \uparrow e}^\dagger}
\newcommand\crbhup[1]{b_{#1, \uparrow h}^\dagger}
\newcommand\beup[1]{b_{#1, \uparrow e}}
\newcommand\bhup[1]{b_{#1, \uparrow h}}

\begin{document}
	
\preprint{}
	
\title{Fractional spin Josephson effect in topological spin superconductors}

\author{Liang Du}
\affiliation{Interdisciplinary Center for Theoretical Physics and Information Sciences, Fudan University, Shanghai 200433, China}

\author{Hua Jiang}
\affiliation{Interdisciplinary Center for Theoretical Physics and Information Sciences, Fudan University, Shanghai 200433, China}
\affiliation{State Key Laboratory of Surface Physics and Institute for Nanoelectronic Devices and Quantum Computing, Fudan University, Shanghai 200433, China}

\author{Yijia Wu}
\thanks{Corresponding author: yijiawu@fudan.edu.cn}
\affiliation{Interdisciplinary Center for Theoretical Physics and Information Sciences, Fudan University, Shanghai 200433, China}
\affiliation{State Key Laboratory of Surface Physics and Institute for Nanoelectronic Devices and Quantum Computing, Fudan University, Shanghai 200433, China}
\affiliation{Hefei National Laboratory, Hefei 230088, China}

\author{X. C. Xie}
\affiliation{Interdisciplinary Center for Theoretical Physics and Information Sciences, Fudan University, Shanghai 200433, China}
\affiliation{Hefei National Laboratory, Hefei 230088, China}
\affiliation{International Center for Quantum Materials, School of Physics, Peking University, Beijing 100871, China}

\date{\today}
\begin{abstract}
Topological spin superconductors are $p$-wave spin-triplet exciton insulators whose topological edge modes have been shown to obey non-Abelian braiding statistics. Based on a toy model as the spin counterpart of the Kitaev's chain, we study the spin Josephson effect adopting the $S$-matrix as well as the Green's function method. The on-site energies of these topological edge modes lead to a transition between the fractional and integer spin Josephson effects. Moreover, non-vanishing on-site energies will also induce a charge pump through the spin Josephson junction. These two effects, distinct features of topological spin superconductors and absent in Majorana systems, can be utilized for spin transport detection of topological spin superconductors.
\end{abstract}
\maketitle


\section{INTRODUCTION} \label{sec:intro}
Exciton insulator (EI) \cite{EI_Mott, EI1, EI2, EI3, EI4, EI5} as the condensate of stable electron-hole pairs has been attracting extensive attention since it was first proposed in the 1960s. Introducing non-trivial band topology into exciton condensates \cite{EI_topological_insulator_film, RuiWang_DYXing_topoEI, XZlLi_topoEI, TMD_MacDonald, Ashivi_Moire, TingxinLi_QAH_Moire} offers an ideal platform for studying the interplay between correlation and topology. In the recent decade, a variety of novel phenomena including the helical-like edge channels \cite{Helical_Edge_Transport_InAs_GaSb}, the quantized edge conductance \cite{RRDu_topoEI_exp}, and the evolution from the helical-like to chiral-like edge transport \cite{wang2023excitonic, pan2023excitonic} have been experimentally observed, suggesting the presence of excitonic topological order. Proposed platforms for realizing topological EIs include the surface of topological insulators \cite{EI_topological_insulator_film, RuiWang_DYXing_topoEI, XZlLi_topoEI}, kagome lattice \cite{FengLiu_flatband, spin_triplet_EI_FengLiu, FengLiu_Kagome}, and Moir\'{e} band systems constructed by bilayer transition metal dichalcogenide (TMD) materials \cite{TMD_MacDonald, Ashivi_Moire, TingxinLi_QAH_Moire}. 


Although the exciton is charge neutral (carries zero electric monopole), it generally carries non-zero higher-order electromagnetic multipole. For instance, the exciton condensate in the electron-hole bilayer possesses non-vanishing electric dipole \cite{eh_bilayer_Eisenstein_MacDonald, EI_topological_insulator_film, EI_review_Eisenstein}. Besides, the Zeeman splitting applied to stabilize the excitons \cite{SSC1, SSC2, Yuanchang_Li_Spin_Triplet_EI, spin_triplet_EI_FengLiu, ShuaiLi_SSC_ABCA_graphene} can lead to spin-polarized excitons. Consequently, the dissipationless flow of spin-triplet excitons forms a super spin current. This exotic state is known as the spin superconductor (SSC) \cite{SSC1, SSC2, ZQBao_GLtheory}, provided that the super spin current screens the spatial gradient of the electric field (known as the electric Meissner effect) \cite{SSC1}. The SSCs support a range of new phenomena such as the spin superfluid \cite{SpinSuperfluidity_nu_0_grapheneQAH, WeiHan_SSC_exp_1} and the spin Josephson effect \cite{SSC1}. Further introducing non-trivial band topology into the SSC leads to even more intriguing states. When $p$-wave electron-hole pairing dominates in a SSC, its topological edge modes have been shown to obey non-Abelian braiding statistics \cite{TSSC_1, review_NSR} due to the non-Abelian geometric phase accumulated via the Aharonov-Casher (AC) effect \cite{ACeffect} on the spin. This is distinctly different from the non-Abelian geometric phase of Majorana zero modes (MZMs) \cite{kitaevchain, Ivanov2001} accumulated by the Aharonov-Bohm effect on the electric charge.


The topological SSCs have been predicted \cite{TSSC_1} to appear in ferromagnetic honeycomb lattice with staggered potential presented \cite{Yuanchang_Li_Spin_Triplet_EI}. However, how to experimentally identify these novel non-Abelian topological edge modes remain an open question. It is known that the electron-hole pairing in EI is reminiscent of the electron-electron pairing in charge superconductor. A feasible method for detecting MZMs as topological edge modes in topological superconductors could be through the fractional Josephson effect \cite{Fu_Kane_Majorana_Joesphson, fractional_Josephson_exp_1, ZMLiao_fractional_Josephson}. The fractional frequency arises from the fact that the charge of both the electron and hole components of the MZM is half that of a Cooper pair. By analogy, the topological edge modes in a topological SSC carry ``half'' the spin compared with the spin-triplet excitons in the bulk states \cite{TSSC_1}. In this way, the spin Josephson effect \cite{SSC1} induced by the topological edge modes in topological SSCs shall exhibit a fractional one-half frequency, making it a potential method for detecting these novel non-Abelian topological edge modes.

In this work, we investigate the fractional Josephson effect in the topological SSCs. The presence of the fractional spin Josephson effect is confirmed by both analytical and numerical evaluations. Moreover, unlike MZMs, the on-site energy for the topological edge modes in the topological SSCs is generally non-zero. This non-zero on-site energy leads to a transition from fractional spin Josephson effect to integer spin Josephson effect. 
Furthermore, adiabatically tuning both the on-site energy and the spin superconducting pairing phase simultaneously can induce a charge pump through the spin Josephson junction. These two effects, which are the distinct features of topological SSC, may facilitate the experimental detection of these non-Abelian topological edge modes in topological SSCs.

The rest of this article is organized as follows. In Sec. \ref{sec:model}, in order to essentially depict the topological SSC, we introduce a toy model as the spin counterpart of the celebrated Kitaev's chain model, which we refer to as the ``spin Kitaev's chain''; In Sec. \ref{sec:spin_jose}, based on this spin Kitaev's chain model, we investigate the fractional spin Josephson effect by adopting the $S$-matrix as well as the Green's function method. After that, in Sec. \ref{sec:charge_pump}, we 
show that a non-zero on-site energy of the topological edge modes in the topological SSCs could lead to a charge pump through the spin Josephson junction. Finally, in Sec. \ref{sec:conclusion}, a conclusion is presented.


\section{Spin Kitaev's chain and its edge modes} \label{sec:model}
As discussed in Sec. \ref{sec:intro}, the topological SSCs may be realized in staggered ferromagnetic graphene or twisted bilayer TMD materials. In order to essentially describe the band topology for the topologically nontrivial SSCs, we introduce a low-energy effective model, which serves as a prototype for the topological SSC model throughout this paper. We start with a spinful one-dimensional chain with a non-interacting Hamiltonian $H_0^{'}$ as

\begin{multline}
H_0^{'} = \sum_{\alpha=\pm} \sum_{\sigma=\uparrow, \downarrow} \alpha \left[ \sum_{x=1}^{N} (E_p+\sigma M^\prime)c^\dagger_{x,\sigma\alpha}c_{x,\sigma\alpha} \right. \\
\left. + \sum_{x=1}^{N-1} (t_0c^\dagger_{x+1,\sigma\alpha}c_{x,\sigma\alpha} + \mathrm{h.c.}) \right]
\label{Ham_nonint}
\end{multline}	
		
\noindent Here, $\alpha=\pm$ are the band indices for the conductance band and the valence band, $\sigma=\uparrow$, $\downarrow$ are the spin indices, $c^\dagger_{x,\sigma\alpha}$ ($c_{x,\sigma\alpha}$) is the electron creation (annihilation) operator on the $x$-th site. $E_p$ is the energy difference between the bottom of the conductance band and the top of the valence band, and $t_0$ represents the nearest neighbour's (NN) electron hopping. Considering the case that the Zeeman splitting term $M'$ applied along the $z$-direction is large enough, we can focus on the lowest-two bands with the indices $\alpha=+$, $\sigma=\uparrow$ and $\alpha=-$, $\sigma=\downarrow$. By performing an electron-hole transformation $\beup{x}=c_{x,\uparrow+}$ and $\crbhup{x}=c_{x,\downarrow-}$ \cite{SSC1, SSC2} in which $\beup{x}$ ($\bhup{x}$) is the annihilation operator for spin up electron (spin up hole), the physics in the vicinity of the Fermi level can be effectively described by a two-band (a spin-up electron band and a spin-up hole band) model

\begin{equation}
\begin{split}
H_0 &= -\sum_{x=1}^{N} M\crbeup{x}\beup{x}+\sum_{x=1}^{N-1}(t_0\crbeup{x+1}\beup{x} + \mathrm{h.c.}) \\
 &- \sum_{x=1}^{N}M\crbhup{x}\bhup{x}+\sum_{x=1}^{N-1}(t_0\crbhup{x+1}\bhup{x} + \mathrm{h.c.})
\end{split}
\label{Ham_nonint_eband_hband}
\end{equation}


\begin{figure}
\centering
\subfigure[]{
\begin{minipage}[t]{7cm}
\centering
\includegraphics[width=6.5cm]{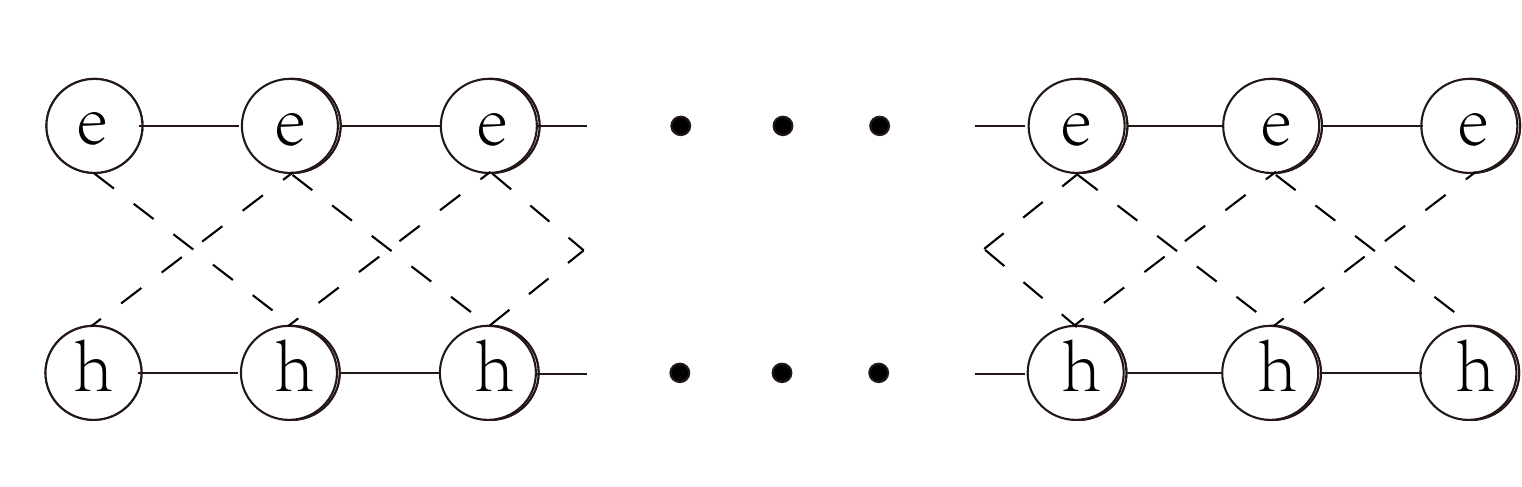}
\end{minipage}
\label{fig:spinkitaechain}
}
\subfigure[]{
\begin{minipage}[t]{8.63cm}
\centering
\includegraphics[width=6.5cm]{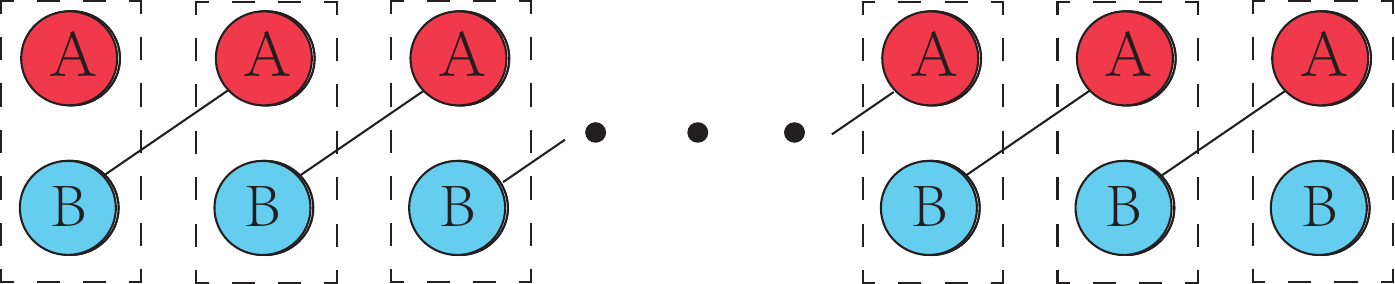}
\end{minipage}
\label{fig:edgestate}
}
\caption{(a) Spin Kitaev's chain composed of a spin up electron chain and a spin up hole chains with NN Coulomb interaction. (b) When $M=0$ and $t_0=|\Delta|$, the edge modes are localized at the ends of the spin Kitaev's chain.}
\end{figure}

\noindent where $M \equiv E_p+M^\prime$ has been defined. In addition to the non-interacting Hamiltonian, the NN Coulomb attraction \cite{swave_plus_pwave_Uchoa_PRL, TSSC_1} between the electron band and the hole band can be introduced as
 
\begin{equation}
H_I = \sum_{x=1}^{N-1}(\Delta\crbeup{x+1}\crbhup{x}+\Delta\crbhup{x+1}\crbeup{x} + \mathrm{h.c.})
\label{Ham_int}
\end{equation}
	 
\noindent where $\Delta \equiv |\Delta| e^{-i\phi} = \langle\bhup{x+1}\beup{x}\rangle = \langle\beup{x+1}\bhup{x}\rangle$ is the NN electron-hole pairing potential after mean-field approximation so that $\phi$ is the spin superconducting pairing phase. 

Combining the non-interacting Hamiltonian with the interacting Hamiltonian as $H = H_0 + H_I$ gives rise to the counterpart of the Kitaev's chain \cite{kitaevchain} in the spin space, so we refer to this model as the spin Kitaev's chain. Performing the Fourier transformations $\beup{k} = \frac{1}{\sqrt{N}}\sum_{x=1}^{N}\beup{x}e^{ikx}$ and $\bhup{k} = \frac{1}{\sqrt{N}}\sum_{x=1}^{N}\bhup{x}e^{-ikx}$, and introducing a 
new basis \cite{SSC1} formed of spin-up electron and spin-up hole as $\begin{pmatrix}
\beup{k}, &\crbhup{k}
\end{pmatrix}^\mathrm{T}$, then the Hamiltonian can be written as 

\begin{equation}
H = \sum_{k}{
\begin{pmatrix}
\crbeup{k}&\bhup{k}
\end{pmatrix} 
H_k 
\begin{pmatrix}
\beup{k} \\ \crbhup{k}
\end{pmatrix}}
\label{Ham_BdG}
\end{equation}

\noindent in which 

\begin{equation}
\begin{split}
H_k = & 2|\Delta|\sin\phi\sin k\sigma_{x} - 2|\Delta|\cos\phi\sin k\sigma_{y} \\
& \qquad +\left(2t_{0}\cos k-M\right)\sigma_{z}
\end{split}
\label{Hk}
\end{equation}

\noindent and $\sigma_{0,x,y,z}$ are Pauli matrices in the basis defined above
. The form of $H_k$ is exactly the same as that used to describe topological superconductors or topological insulators. Therefore, in analogy with the topological insulators, a Zak phase $\gamma$ \cite{Zakphase1, Zakphase2} can also be defined as the topological invariant describing this spin Kitaev's chain model ($\gamma=0$ for trivial phase $|M|>2|t_0|$, and $\gamma=\pi$ for topological phase $|M|<2|t_0|$). The difference lies in the physical interpretation of the Hamiltonian basis. For one-dimensional topological superconductors depicted in the Bogoliubov-de Gennes (BdG) basis, the topological edge modes are MZMs. By contrast, the topological edge modes in one-dimensional topological insulators are Dirac fermionic modes well known as Jackiw-Rebbi zero modes \cite{JackiwRebbi}. In the spin Kitaev's chain, 
the topological edge modes are also Dirac fermionic modes, but with a specific spin structure. For demonstrating the properties of the edge modes in topological SSC, we choose a special set of parameters $M=0$ and $t_0=|\Delta|$ in the topological phase of the spin Kitaev's chain so that the Hamiltonian is reduced to
	 
\begin{equation}
H = 2|\Delta|\sum_{x=1}^{N-1}\Psi^\dagger_{x+1,\mathrm{A}}\Psi_{x,\mathrm{B}} + \mathrm{h.c.}
\label{Ham_basisAB}
\end{equation}

\noindent Here, a set of new operators

\begin{equation}
\Psi_{x,\mathrm{A}(\mathrm{B})} \equiv \frac{1}{\sqrt{2}}(e^{i\phi/2}\beup{x} \mp e^{-i\phi/2}\crbhup{x})
\label{new_operators}
\end{equation}

\noindent have been defined, in which the $\mp$ corresponds to the $\Psi_{x,\mathrm{A}}$ and $\Psi_{x,\mathrm{B}}$, respectively. Equation (\ref{Ham_basisAB}) indicates that there are two isolated zero-energy modes $\Psi_{1,\mathrm{A}}$ and $\Psi_{N,\mathrm{B}}$ perfectly-localized at the ends of the spin Kitaev's chain [see Fig. \ref{fig:edgestate}]. It is easy to show that they are not self-conjugate and satisfy anticommutation relations $\{\Psi_{x,\mathrm{A}},\Psi^\dagger_{x',\mathrm{A}}\}=\{\Psi_{x,\mathrm{B}},\Psi^\dagger_{x',\mathrm{B}}\}=\delta(x-x')$. Therefore, these two topological edge modes are Dirac fermion zero modes other than the self-conjugate Majorana ones. 

The bulk states in topological SSCs are spin-triplet excitons $\crbeup{x}\crbhup{x}$ carrying spin $2(\hbar/2)$ and zero charge $q=0$. In contrast, the edge modes $\Psi_{1,\mathrm{A}}$ and $\Psi_{N,\mathrm{B}}$ are constructed by the superposition of the annihilation operator of spin-up electron and the creation operator of spin-up hole. Therefore, these edge modes carry non-zero charge $q=e$, while their spin components along the $z$-direction vanish as $(-\hbar/2)+(\hbar/2)=0$. Although being spin non-polarized along the $z$-direction, both the $(-\hbar/2)$-component $\beup{x}$ and the $(\hbar/2)$-component $\crbhup{x}$ in Eq. (\ref{new_operators}) could acquire an AC phase via coupling to the magnetic field. To be specific, when the spin superconducting pairing phase $\phi$ in Eq. (\ref{Ham_int}) changes by $2\pi$, these topological edge modes [Eq. (\ref{new_operators})] will obtain an extra factor as $e^{i\phi/2}=e^{-i\phi/2}=e^{i\pi}$. This $e^{i\pi}$ factor is precisely the source of the non-Abelian braiding statistics \cite{TSSC_1} exhibited by these topological edge modes in topological SSCs.

 
When two SSCs are weakly coupled and form a spin Josephson junction, a spin current driven by the spin superconducting pairing phase difference $\phi \equiv \phi_L - \phi_R$ will flow through this junction. Such an effect is known as the spin Josephson effect \cite{SSC1, ZQBao_GLtheory}. As demonstrated above, the wavefunctions of the edge modes in topological SSC come back to their original forms only when the superconducting pairing phase $\phi$ changes by $4\pi$ (other than $2\pi$). Hence, these topological edge modes can give rise to a one-half fractional frequency (i.e., $4\pi$ period) spin Josephson effect with respect to $\phi$. Experimentally, $\phi$ can be tuned through the difference of the Zeeman splitting strength between the two SSCs in the spin Josephson junction. In the following section, based on the spin Kitaev's chain model, we investigate the fractional spin Josephson effect by adopting the $S$-matrix as well as the Green's function method.


\section{Fractional spin Josephson effect} \label{sec:spin_jose}

\subsection{$S$-MATRIX} \label{sec:Smatrix}
By utilizing the $S$-matrix method, one can conveniently describe the transport induced by the low-lying edge modes near the Fermi level. Notably, in contrast to MZMs whose zero-energy property is protected by the particle-hole symmetry \cite{zeroenergy}, the edge modes in topological SSCs, being Dirac fermionic modes, can possess an on-site energy (For instance, when an electric gate voltage is imposed). Therefore, the BdG-like equation describing the SSC with an on-site energy applied is

\begin{equation}
	\begin{pmatrix}
		\xi_k-M-\mu_{\beta}+U(x) & \Delta_{\beta\gamma}\\
		\Delta^\dagger_{\beta\gamma} & -\xi_k+M-\mu_{\beta}+U(x)
	\end{pmatrix} \Psi = E\Psi
	\label{BdG_equiation_Smatrix}
\end{equation}

\noindent Here, $\xi_k=-i\hbar v_F\partial_x$ is the kinetic energy, $v_F=\hbar k_F/m$ is the Fermi velocity, 
and $U(x)=U_0\delta(x)$ depicts the potential barrier between the left and the right SSC in the spin Josephson junction. The on-site energy applied on the left/right SSC is denoted as $\mu_{\beta}$ with $\beta = L, R$. Without loss of generality, we set $\mu_L = \mu_0+\mu$ and $\mu_R = \mu_0$, assuming $\mu \ll \mu_0$ in the following discussion.

\begin{figure}
	\centering
	\subfigure[]{
		\begin{minipage}[t]{8cm}
			\centering
			\includegraphics[width=8cm]{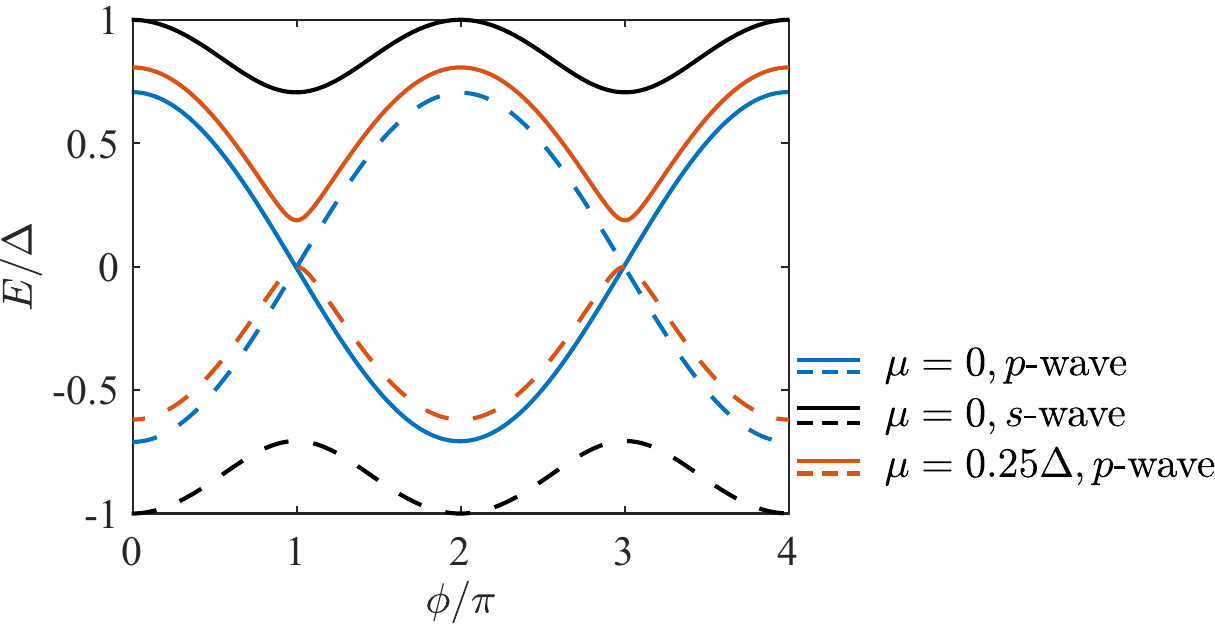}
		\end{minipage}
		\label{fig:S_spectrum}
	}
	\centering
	\subfigure[]{
		\begin{minipage}[t]{8cm}
			\centering
			\includegraphics[width=8cm]{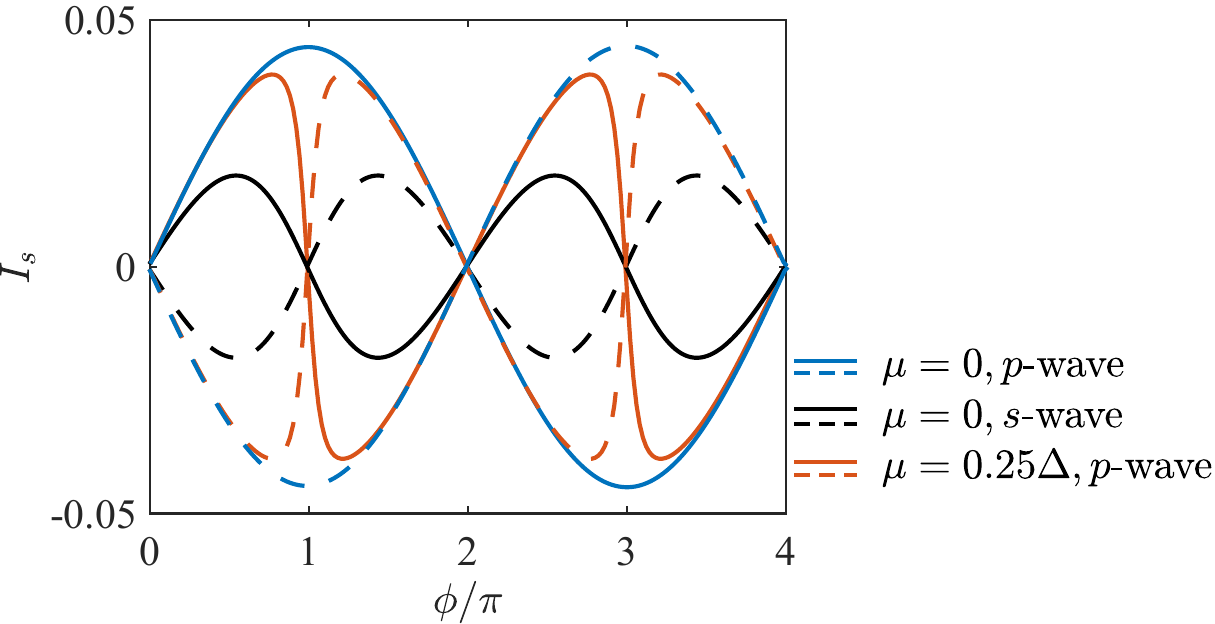}
		\end{minipage}
		\label{fig:S_jos}
	}
	\caption{(a) Energy spectrum of the low-lying modes near the interface of the spin Josephson junction with respect to the phase difference $\phi$. (b) Josephson spin current $I_s=-\partial E/\partial\phi$ with respect to the phase difference $\phi$.
	Blue line: $p$-wave spin Josephson junction with $\mu=0$; 
	Black line: $s$-wave spin Josephson junction  with $\mu=0$; 
	Red line: $p$-wave spin Josephson junction with $\mu=0.25\Delta$. 
		Dashed (solid) lines correspond to the $\pm$ in Eqs. (\ref{p_wave_spectrum}) and (\ref{s_wave_spectrum}). Other parameters chosen are $D=0.5$, and $\Delta=2$.}
\label{fig:S_spectrum_and_S_jos}
\end{figure}

In the case of $p$-wave pairing, the spin superconducting pairing potential is in the form of $\Delta_{\beta\gamma} = \langle b_{k, \uparrow \gamma} b_{k, \uparrow \bar{\gamma}} \rangle = \Delta_\beta k_x/k_F$ where $\hbar k_F=\sqrt{2m\mu_0}$. Here, $\beta = L, R$ is the index for the left/right SSC, and $\gamma=\pm$ is the electron/hole index ($\gamma=+$ for $\langle b_{k, \uparrow e} b_{k, \uparrow h} \rangle$, and $\gamma=-$ for $\langle b_{k, \uparrow h} b_{k, \uparrow e} \rangle$). We assume $\Delta_L = \Delta$ and $\Delta_R = \Delta e^{-i\phi}$ so that $\phi$ is the phase difference in this spin Josephson junction. By imposing the boundary condition at $x=0$, the energy spectrum of the low-lying modes bound at the interface of the spin Josephson junction constructed by $p$-wave SSCs is \cite{kwon}

\begin{multline}
E^{p\mbox{-}\mathrm{wave}}_{\pm}(\phi) = \frac{2-D}{4}\mu \\
\pm \sqrt{\left(\frac{2-D}{4}\right)^2 \mu^2 + D\Delta^2\cos^2\left(\frac{\phi}{2}\right)}
\label{p_wave_spectrum}
\end{multline}

\noindent where $D \equiv \frac{1}{(mU_0/\hbar^2k_F)^2+1}$ is the transmission coefficient. As shown in Fig. \ref{fig:S_spectrum}, this spectrum is in a $4\pi$ period ($2\pi$ period) with respect to the spin superconducting pairing phase difference $\phi$ when the on-site energy difference $\mu=0$ ($\mu \neq 0$). Since the spin Josephson current $I_s$ can be obtained via the derivative \cite{spinphase} of $E^{p\mbox{-}\mathrm{wave}}_{\pm}(\phi)$ with respect to $\phi$, the spin Josephson current also exhibits a $4\pi$ periodicity in $\phi$, a phenomenon known as the fractional spin Josephson effect, when the on-site energy difference is zero. In contrast, as shown in Fig. \ref{fig:S_jos}, a non-zero on-site energy difference $\mu \neq 0$ lead to a transition, restoring the fractional spin Josephson effect to an integer spin Josephson effect.

For the sake of comparison, the spin Josephson effect exhibited by the topologically trivial $s$-wave pairing SSCs is also shown in Fig. \ref{fig:S_spectrum_and_S_jos}. For the spin Josephson junction constructed by $s$-wave SSCs, the spin superconducting pairing is described by $\Delta_{\beta\gamma} = \langle b_{k, \uparrow \gamma} b_{k, \uparrow \bar{\gamma}} \rangle = \gamma\Delta_\beta$, and the energy spectrum of the low-lying modes can still be determined by $S$-matrix as \cite{kwon}

\begin{equation}
E^{s\mbox{-}\mathrm{wave}}_{\pm}(\phi) = \frac{D}{4}\mu \pm \sqrt{\left(\frac{D}{4}\right)^2 \mu^2 + \Delta^2 \left[1-D\sin^2(\phi/2) \right]}
\label{s_wave_spectrum}
\end{equation}

\noindent It indicates that the spin Josephson current induced by topologically trivial $s$-wave SSC consistently exhibits an integer spin Josephson effect with a $2\pi$ period, regardless of whether the on-site energy difference $\mu = 0$ or $\mu \neq 0$.


\subsection{Green's function method} \label{sec:GF}
To account for the contributions from both the bulk states and the edge modes, we adopt the Green's function method, which is closer to realistic experimental device and has been widely used to investigate the fractional Josephson effect in topological superconductors \cite{GF1, GF2, GF3, trifractional, JJQi_Review}. A spin Josephson junction can be described by lattice model as $H = \sum_{\beta=L,R} H_{\beta} + H_t$, where $H_\beta$ is the Hamiltonian for the left/right SSC ($\beta=L,R$). 
As has been stated above, the spin Josephson junction can generally possess non-zero on-site energy term. Therefore, we choose $H_L = H(\phi) + \mu\sigma_0$ and $H_R = H(\phi=0) - \mu\sigma_0$ so that the different chemical potentials $\pm\mu$ result in an on-site energy difference for the topological edge modes in the left and right SSCs. Here, $H(\phi)$ represents the real space Hamiltonian of spin Kitaev's chain (see Sec. \ref{sec:model}) with a chain length $N$ and a spin superconducting pairing phase $\phi$.
Besides, the tunneling Hamiltonian $H_t$ between these two SSCs reads

\begin{equation}
H_t = \varPhi^\dagger_{L} \hat{t}_{\mathrm{LR}} \varPhi_{R} + \mathrm{h.c.}\label{Ham_tunnel}
\end{equation}

\noindent Here, the basis chosen is $\varPhi_{R} \equiv \begin{pmatrix}
\beup{1\mathrm{R}}, &\crbhup{1\mathrm{R}} \end{pmatrix}^\mathrm{T}$, and $\varPhi_{L} \equiv \begin{pmatrix} \beup{N\mathrm{L}}, &\crbhup{N\mathrm{L}} \end{pmatrix}^T$, in which the subscript ``$1\mathrm{R}$'' and ``$N\mathrm{L}$'' denotes the first site in the right SSC and the $N$-th site (the last site) in the left SSC, respectively. $\hat{t}_{\mathrm{LR}} \equiv t_{\mathrm{LR}} \sigma_z$ describes the coupling between these two topological SSCs. Utilizing the Heisenberg equation of motion, the spin current flowing through this spin Josephson junction can be calculated via

\begin{equation}
\begin{split}
I_s &= \frac{\hbar}{2}(I_\uparrow-I_\downarrow) = \frac{\hbar}{2}\frac{\partial}{\partial t}(\langle\hat{N}_\uparrow\rangle - \langle\hat{N}_\downarrow\rangle) \\
&= -\frac{1}{4\pi} \int{\mathrm{d}\omega \mathrm{Tr}(\sigma_z G^<_{\mathrm{RL}} \hat{t}_{\mathrm{LR}} + \mathrm{h.c.})}
\end{split}
\label{GF_spin_current}
\end{equation}

\noindent In the same way, the charge current flowing through this spin Josephson junction can also be evaluated by Green's function as $I_c = -e(I_\uparrow+I_\downarrow) = \frac{e}{h} \int{\mathrm{d}\omega \mathrm{Tr}(G^<_{\mathrm{RL}} \hat{t}_{\mathrm{LR}} + \mathrm{h.c.})}$. Since the left and the right SSCs are directly coupled here, the Green's functions can be evaluated via the Dyson's equations $G^r_{\mathrm{LL}}=g^r_{\mathrm{LL}}+g^r_{\mathrm{LL}} \hat{t}_{\mathrm{LR}} G^r_{RL}$, and $G^r_{\mathrm{RL}} = g^r_{\mathrm{RR}} \hat{t}_{\mathrm{RL}} G^r_{\mathrm{LL}}$. Here, $g^{r}$, $g^{a}$, and $g^{<}$ are the surface retarded, advanced, and distribution Green's functions in the absence of the tunneling Hamiltonian $H_t$, respectively. The distribution Green's function satisfies $g^< = f(\epsilon)(g^a-g^r)$ where $f(\epsilon)$ is the Fermi distribution function and $g^a = (g^r)^\dagger$. In this paper, we consider zero temperature only. 
Noticing again that this spin Josephson current is driven by the spin superconducting phase difference $\phi$. 
Experimentally, the ``spin voltage bias'' driving the spin current can be implemented by introducing a difference in the magnetic field (i.e., a difference in the Zeeman splitting) applied to the left and right SSCs.
	             
\begin{figure}
\centering
\subfigure[]{
\begin{minipage}[t]{8cm}
\centering
\includegraphics[width=8cm]{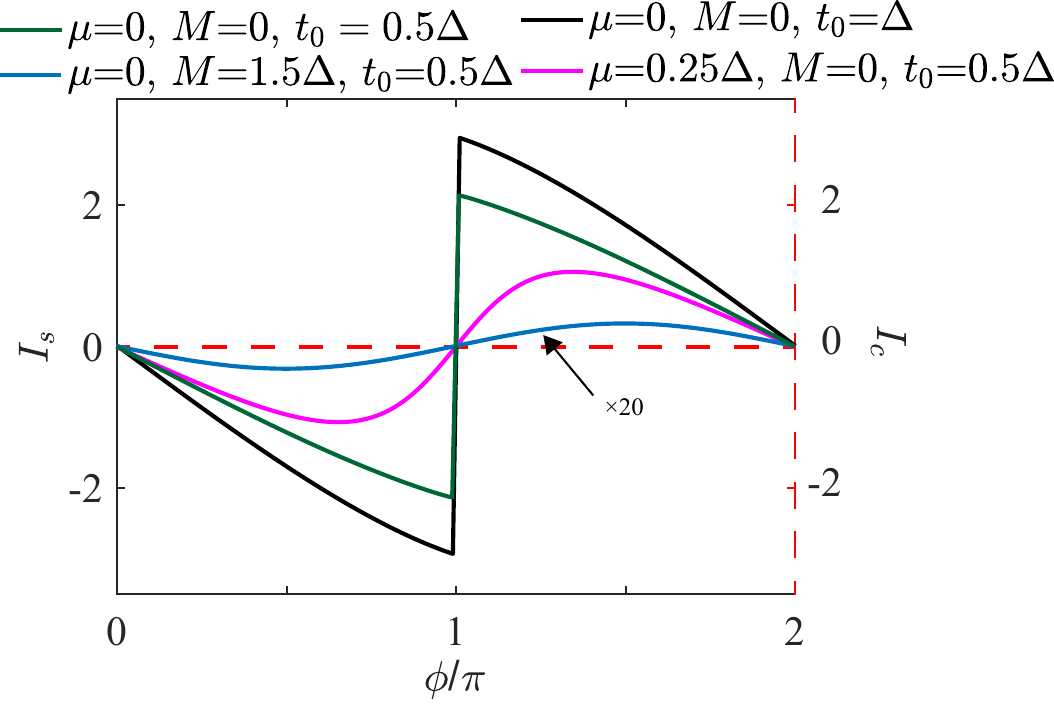}
\end{minipage}
\label{fig:pjos}
}
\subfigure[]{
\begin{minipage}[t]{8cm}
\centering
\includegraphics[height=4cm]{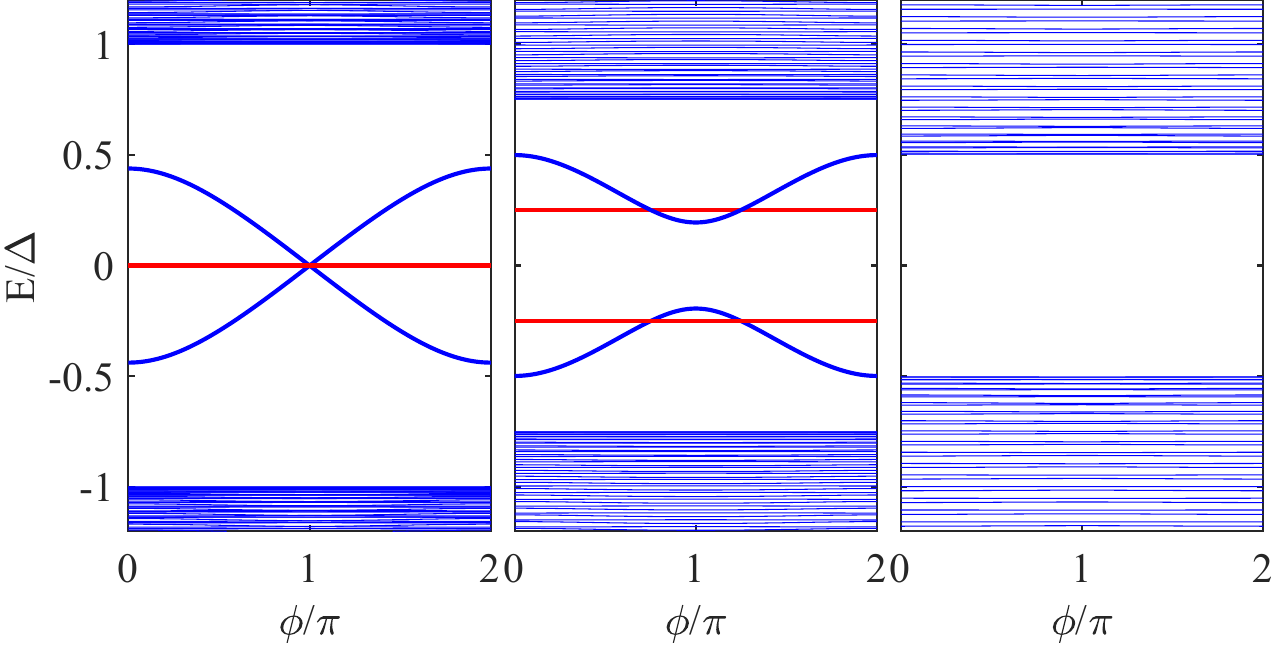}
\end{minipage} 	  	   		
\label{fig:p-spectrum}
}
\caption{(a) The spin Josephson current $I_s$ with respect to the phase difference $\phi$, which is obtained by the Green's function method. 
Green: $\mu=0$, $M=0$, and $t_0=0.5\Delta$; 
Black: $\mu=0$, $M=0$, and $t_0=\Delta$; 
Blue: $\mu=0$, and $M=3t_0=1.5\Delta$; 
Magenta: $\mu=0.25\Delta$, $M=0$, and $t_0=0.5\Delta$. 
Red dashed line: the charge current $I_c$ vanishes for all four sets of parameters mentioned above. The spin Josephson current $I_s$ represented by the blue line is magnified $20$ times for clarity. 
(b) Energy spectrum for both the bulk states and edge modes. 
Left: $\mu=0$, $M=0$, and $t_0=0.5\Delta$ [corresponds to the green line in (a)]; 
Middle: $\mu=0.25\Delta$, $M=0$, and $t_0=0.5\Delta$ [corresponds to the magenta line in (a)]; 
Right: $\mu=0$, $M=1.5\Delta$, and $t_0=0.5\Delta$ [corresponds to the blue line in (a)]. Other parameters chosen are $t_{\mathrm{LR}}=0.5\Delta$, and both the left and right SSC chains consist of $N=100$ sites.}
\end{figure}

The surface Green's functions $g^{r}$, $g^{a}$, and $g^{<}$ above can be numerically investigated by iteration \cite{diedai, diedai2}. 
The spin Josephson current (including both the contributions from the bulk states and the edge modes) obtained by Green's function is shown in Fig. \ref{fig:pjos}, and the corresponding energy spectrum of the spin Josephson junction  is shown in Fig. \ref{fig:p-spectrum}. 
For a topologically non-trivial $p$-wave spin Josephson junction ($|M| < 2|t_0|$) with zero on-site energy difference ($\mu=0$), there are two subgap edge modes at the outermost ends of the spin Josephson junction whose energies remain invariant with respect to $\phi$ [see the red lines in the left column of Fig. \ref{fig:p-spectrum}]. In addition to this, the finite-size-effect-induced coupling \cite{finitesize} of the two topological edge modes near the interface of the junction leads to a pair of symmetric/antisymmetric states \cite{trifractional}. As a consequence, the energies of two subgap states cross at $\phi=\pi$ [see the left column of Fig. \ref{fig:p-spectrum}], which leads to a sudden jump of the spin Josephson current [see the black and green lines in Fig. \ref{fig:pjos}]. Such a behaviour can also be interpreted in the scheme of $S$-matrix, as the eigenenergies formed by the topological edge modes cross at $\phi=\pi$ [see the solid blue line and the dashed blue line in Fig. \ref{fig:S_spectrum}], which leads to a sudden jump of the spin Josephson current from the dashed blue line in Fig. \ref{fig:S_jos} to the solid blue line in Fig. \ref{fig:S_jos} at $\phi=\pi$. 
In this manner, at zero temperature and in the DC limit \cite{YHLi_SCPMA}, the $4\pi$ period spin Josephson effect manifests as a $2\pi$-period spin Josephson current with an abrupt jump in the spin current at $\phi=\pi$. At the same time, as shown in the red dashed line in Fig. \ref{fig:pjos}, the charge current $I_c$ flowing through this junction vanishes. This is quite reasonable because the contributions from the electron and the hole exactly cancel each other out.

For a topologically non-trivial $p$-wave spin Josephson junction ($|M|<2|t_0|$), when the topological edge modes possess non-vanishing on-site energies ($\mu \neq 0$), the energy spectrum of the two eigenstates formed by the topological edge modes near the interface of the junction no longer crosses at $\phi=\pi$ [see the middle column of Fig. \ref{fig:p-spectrum}]. 
In this way, the sudden jump of the spin Josephson current at $\phi=\pi$ is now also absent [see the magenta line in Fig. \ref{fig:pjos}], and the fractional spin Josephson effect comes back to integer spin Josephson effect with a $2\pi$ period. This behaviour is consistent with the analytic formula [Eq. (\ref{p_wave_spectrum})] obtained from the $S$-matrix method. Finally, for topologically trivial case ($|M| > 2|t_0|$), the subgap topological edge modes are absent [see the right column of Fig. \ref{fig:p-spectrum}]. As a result, the spin Josephson current is entirely induced by the bulk states far away from the Fermi level, and the corresponding spin Josephson current is quite small and exhibits smooth behavior with respect to $\phi$, following a $2\pi$-periodicity as expected [see the blue line in Fig. \ref{fig:pjos}]. 
 The fact that the spin Josephson current in the topologically trivial case is induced by excitonic bulk states is further supported by additional numerical results that the spin Josephson current is proportional to the square of the single-particle tunneling amplitude as $I_s \propto t^2$ (see Appendix \ref{appendix:relation}). As a comparison, when the spin Josephson current is induced by the topological edge modes, it is directly proportional to the single-particle tunneling amplitude as $I_s \propto t$ (see Appendix \ref{appendix:relation}).

 It is worth noticing that such a sudden jump of the spin Josephson current can also be observed in a topologically trivial $s$-wave spin Josephson junction, provided a localized spin-$1/2$ state exists in the insulating barrier of this junction. When a spin flip occurs in this localized state, the spin Josephson current similarly exhibits a sudden jump (see Appendix \ref{appendix:QD}). Therefore, to identify the topological edge modes in topological SSCs through the presence of a sudden jump in the spin Josephson current, it is essential to rule out the possibility of a local spin flip within the insulating barrier. 
In addition to the spin Josephson current, as will be shown in the following section, an adiabatic tuning of both the on-site energy and the spin superconducting pairing phase can induce a charge pump through the $p$-wave spin Josephson junction. Since the steady current flow in the spin Josephson effect is charge neutral, this charge pump is more noticeable and can serve as additional evidence for identifying topological SSCs, complementing the fractional spin Josephson effect.


 	  	    
\section{charge pump} \label{sec:charge_pump}

\begin{figure} 
	\centering
	\subfigure[]{
		\begin{minipage}[t]{6.5cm}
			\centering
			\includegraphics[width=5cm]{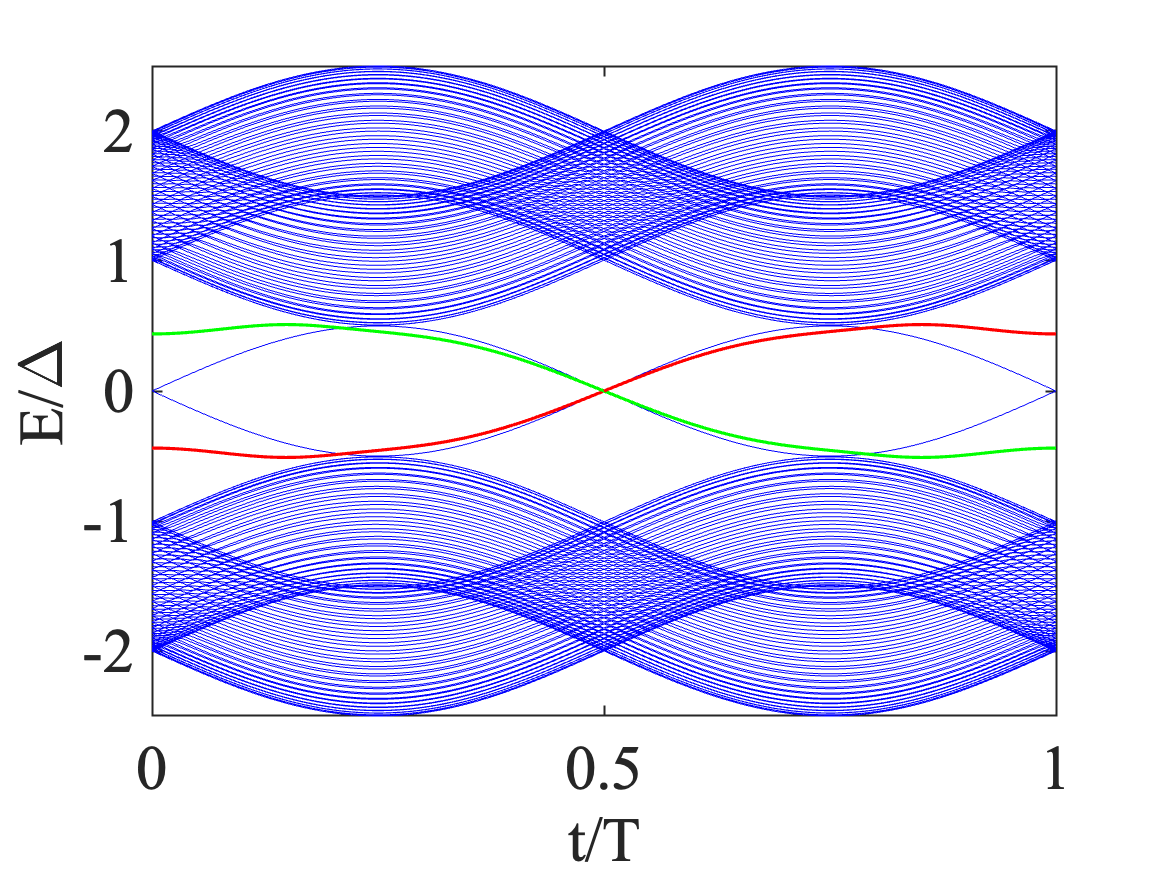}
		\end{minipage}
		\label{fig:pump}
	}
	\subfigure[]{
		\begin{minipage}[t]{6.5cm}
			\centering
			\includegraphics[width=5cm]{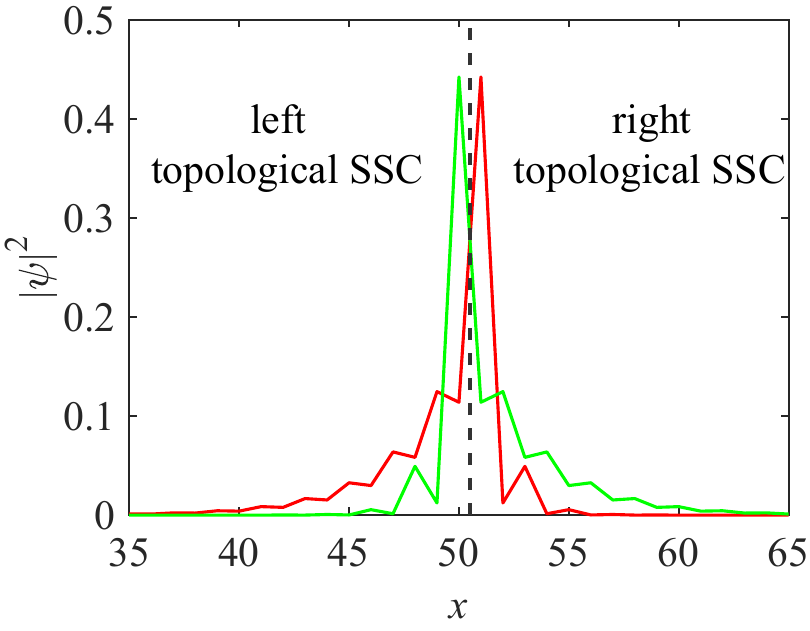}
			\label{fig:distribution}
		\end{minipage}
	}
	\caption{(a) The instantaneous energy spectrum as a function of time $t$ over one period of the charge pump. The parameters chosen here are $M=0$, $t_{\mathrm{LR}} = t_0 = 0.5\Delta$, $\delta\mu_{0} = 0.5\Delta$ so that the time-dependent pumping parameters are $\mu_L = 0.5\Delta \sin\left(\frac{2\pi t}{T}\right)$, $\mu_R = -0.5\Delta \sin\left(\frac{2\pi t}{T}\right)$, and the phase difference $\phi = 2\pi t/T$. Both the left and the right topological SSC chains consist of $N=50$ sites. (b) Wavefunction distribution of the edge modes highlighted in red/green in (a) at $t=0.25T$ during the pump, at this point $\mu_L = 0.5\Delta$, $\mu_R = -0.5\Delta$, and $\phi = \pi/2$.}
\end{figure}

In addition to causing a transition from the fractional to the integer spin Josephson effect, an on-site energy difference between the left/right topological SSCs could also lead to a charge pump in a $p$-wave spin Josephson junction. This is reminiscent of the Fu-Kane spin pump \cite{spin_pump_1, spin_pump_2} in the Rice-Mele model \cite{RiceMele}, in which an on-site energy difference is introduced into the Su-Schrieffer-Heeger model \cite{SSH_model} describing the one-dimensional topological insulator. For observing this a charge pump, we investigate a $p$-wave spin Josephson junction with time-dependent on-site energy difference 
as well as time-dependent spin superconducting phase difference. 
As an illustration, the on-site energies chosen for the left/right SSCs are $\mu_L = +\mu(t)$, and $\mu_R = -\mu(t)$, respectively, where $\mu(t) = \delta\mu_{0} \sin\left(\frac{2\pi t}{T}\right)$. The phase difference $\phi(t) = 2\pi t/T$ that $T$ is the time-cost for one period of pump satisfying the adiabatic condition. As illustrated in Fig. \ref{fig:pump}, numerical calculation shows that there are two subgap modes whose energies vary with the on-site energies as sinusoidal functions (depicted as the two subgap blue lines), which correspond to the edge modes localized at the outermost ends of the spin Josephson junction. In addition, the energy spectra of two other edge modes (highlighted in red and green), which are localized near the junction's interface, intersect only once during a single period of the charge pump (the energy level crossing at $t=T$ is absent). This behavior indicates that an edge mode (highlighted in red) is adiabatically transported from the valence band to the conductance band. At the same time, another edge mode (highlighted in green) is adiabatically moved from the conductance band to the valence band. This instantaneous energy spectrum reflects the evolution of the edge modes from one side of the interface of the spin Josephson junction to the other side.

Specifically, as shown in Fig. \ref{fig:distribution}, at $t=0.25T$ during the pump, where the on-site energies are $\mu_L = 0.5\Delta$ and $\mu_R = -0.5\Delta$, the wavefunction distributions reveal that the state marked by red (green) corresponds to the edge mode centered at the right (left) side of the spin Josephson junction's interface. At this point, the edge mode on the right side of the interface possesses a lower energy than the one on the left side. Consequently, the edge mode being occupied at this time is situated at the right side of the spin Josephson junction. 
Afterwards, as $t$ increases from $t=0.25T$ to $t=0.75T$, the energy spectra of these two edge modes cross at zero energy when $t=0.5T$. This indicates that at $t=0.5T$, where the on-site energies $\mu_L = \mu_R = 0$ and the phase difference $\phi=\pi$, the spatial overlap of these two edge modes does not lift their degeneracy [see also the energy level crossing shown in the left column of Fig. \ref{fig:p-spectrum}]. In other words, these two edge modes do not couple at $\phi=\pi$. In this way, the system, starting in the ground state at $t=0.25T$, evolves into an excited state at $t=0.75T$ \cite{spin_pump_2}, in which the edge mode with higher energy on the right side of the interface is occupied, while the edge mode with lower energy on the left side is vacant.


The intriguing thing is that at $t=T$, where the on-site energies $\mu_L=\mu_R=0$ and the phase difference $\phi=2\pi$, these two edge modes no longer cross at $E=0$. Notably, the absence of this energy level crossing has already been exhibited in the left column of Fig. \ref{fig:p-spectrum} at $\phi=2\pi$. 
This degeneracy lift is attributed to the spatial-overlap-induced coupling between these two edge modes at $\phi=2\pi$. As a result, the eigenstates at this point are symmetric or antisymmetric combinations of these two edge modes, and the occupied state at $t=T$ is evenly distributed across both the left and right sides of junction's interface. As $t$ increases further to $t=1.25T$ (equivalent to $t=0.25T$), the occupied state becomes the edge mode localized on the left side of the interface [highlighted in green in Fig. \ref{fig:pump}]. Consequently, over one period of the charge pump, from $t=0.25T$ to $t=1.25T$, an edge mode is pumped from the right side of the interface to the left side.

It is evident that the charge pump arises from the fact that the energy spectra of the edge modes cross only once, rather than twice, during one period of the pump. This behavior can be attributed to the $4\pi$-periodicity of the edge modes' wavefunctions [see Eq. (\ref{new_operators})], in which $\Psi_{x,\mathrm{A(B)}}$ changes its form from a symmetric (anti-symmetric) state to an anti-symmetric (symmetric) state as $\phi$ increases from $\phi=\pi$ to $\phi=2\pi$.
At $\phi=2\pi$, the non-zero wavefunction overlap between the edge modes on the left and right sides of the interface results in significant coupling between these modes, leading to a degeneracy lift. In contrast, at $\phi=\pi$,  this wavefunction overlap integral vanishes because the symmetry of the wavefunction changes.  Consequently, the edge modes do not couple at $\phi=\pi$, and any degeneracy lift is prohibited.

As has been discussed in Sec. \ref{sec:model}, the edge modes of topological SSCs carry non-zero charge $q=e$, while their spins are non-polarized along the $z$-direction. Consequently, adiabatically tuning the on-site energy difference and the spin superconducting pairing phase simultaneously result in a charge pump through the spin Josephson junction. 
Since this charge pump arises from the $4\pi$-periodicity of the topological edge modes with respect to $\phi$, it is intrinsically related to the fractional spin Josephson effect. From the perspective of symmetry classification, when the on-site energy is absent ($\mu=0$), the $p$-wave SSC can be classified into symmetry class D \cite{AZ_class, class}, in which the particle-hole symmetry (PHS) operator is defined as $\mathcal{P}=\sigma_{x}\mathcal{K}$
($\mathcal{K}$ is the complex conjugation operator) and the Hamiltonian in Eq. (\ref{Hk}) satisfies $\mathcal{P} H_k \mathcal{P}^{-1} = -H_{-k}$. Teo and Kane have demonstrated that \cite{classApump} only the fermion parity pump is permitted for a topological system in symmetry class D. In contrast, the inclusion of an on-site energy term $\mu\sigma_0$ breaks this PHS and transitions this $p$-wave SSC into symmetry class A, in which the charge pump becomes allowed \cite{classApump}. This indicates that the on-site energy breaking the PHS is indispensable in this charge pump. Consequently, this charge pump is a distinct feature of topological spin superconductors being absent in Majorana systems.


 	  	    
\section{Conclusions} \label{sec:conclusion}
In summary, in this article, we have introduced the spin Kitaev's chain model as a toy model describing the one-dimensional topological SSC. Utilizing the $S$-matrix method as well as the Green's function method, we have shown that the $4\pi$-period fractional spin Josephson effect in the $p$-wave spin Josephson junction manifests itself as a $2\pi$-period spin Josephson current with an abrupt jump. Nonetheless, this kind of behavior can also be observed in a topologically trivial spin Josephson junction, provided that the spin of a localized state in the insulating barrier of the junction flips as the spin superconducting phase difference $\phi$ changes. Hence, the $2\pi$-period spin Josephson current with an abrupt jump of the current alone may not provide exclusive evidence for the presence of topological edge modes in $p$-wave SSCs unless processes that violate parity conservation, such as local spin flips, are ruled out.

Compared with the MZMs whose zero-energy property is protected by PHS, the topological edge modes in $p$-wave SSCs can generally possess non-zero on-site energy. We have shown that a non-vanishing on-site energy difference in the $p$-wave spin Josephson junction can induce a transition from the $4\pi$-period fractional spin Josephson effect to the $2\pi$-period integer spin Josephson effect. Moreover, adiabatically tuning this on-site energy difference as well as the spin superconducting phase difference $\phi$ could induce a charge pump through the spin Josephson junction. 
Notably, this type of charge pump is prohibited in superconducting systems preserving the PHS. Therefore, in addition to the fractional spin Josephson effect, the charge pump induced by the on-site energy can also serve as an experimental probe to detect these topological edge modes that reside in $p$-wave SSCs and obey non-Abelian braiding statistics.


\section{Acknowledgements}
The authors thank Yu-Hang Li for fruitful discussions. This work is financially supported by the National Key R\&D Program of China (Grants No. 2019YFA0308403), the Innovation Program for Quantum Science and Technology (Grant No. 2021ZD0302400), the National Natural Science Foundation of China (Grant No. 12304194), and Shanghai Municipal Science and Technology (Grant No. 24DP2600100).

\appendix

	 
\section{SPIN JOSEPHSON CURRENT AND THE TUNNELING AMPLITUDE}
\label{appendix:relation}
	 
\begin{figure}
\centering
\begin{minipage}[t]{10cm}
\centering
\includegraphics[width=8.5cm]{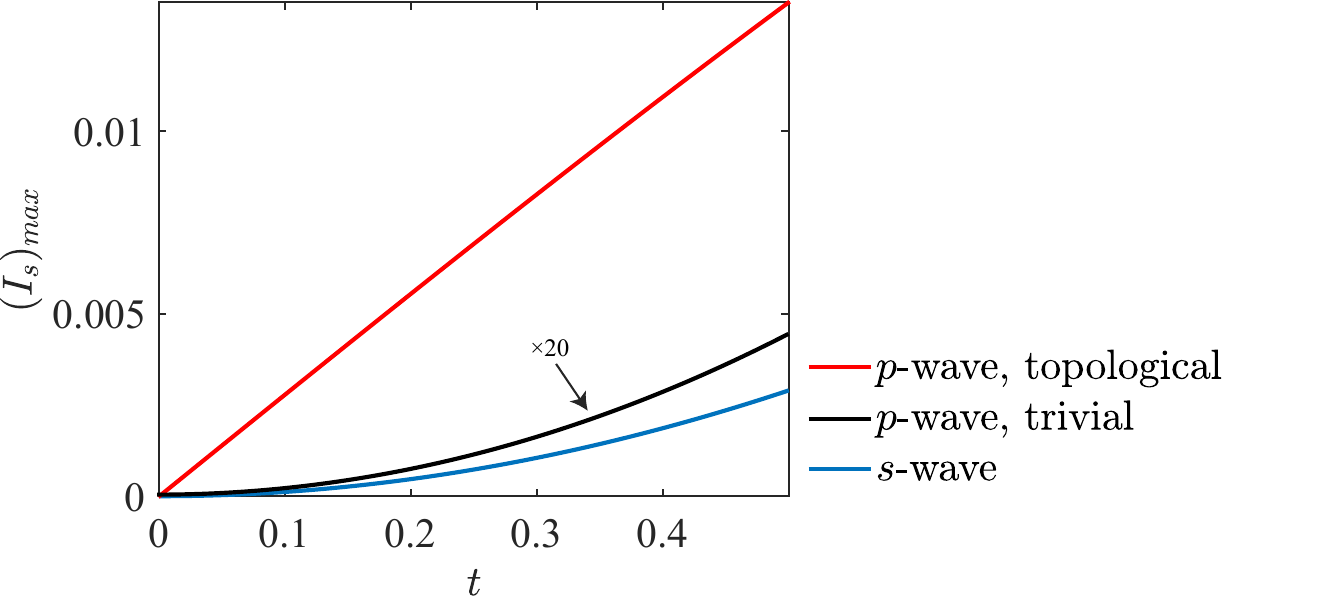}
\end{minipage}
\caption{The relation between the maximum value of the spin Josephson current $(I_s)_{\mathrm{max}}$ and the tunneling amplitude $t$. 
Red line: $p$-wave spin Josephson junction in topological phase ($\mu=M=0$); 
Black line: $p$-wave spin Josephson junction in trivial phase ($\mu=0$, and $M=1.1\Delta$); 
Blue line: $s$-wave spin Josephson junction ($\mu=M=0$). The other parameters are $t_{\mathrm{LR}}=t_0=0.5\Delta$. The $(I_s)_{\mathrm{max}}$ represented by the black line is magnified 20 times for clarity.}
\label{fig:hopping}
\end{figure}

The $S$-matrix method [see Eqs. (\ref{p_wave_spectrum}) and (\ref{s_wave_spectrum})] have shown that when $\mu_L=\mu_R=0$, the spin Josephson current for $p$-wave SSC is proportional to the square root of the transmission coefficient $D$ as $I^{p\mbox{-}\mathrm{wave}}_s \propto \sqrt{D}$. By contrast, the spin Josephson current for $s$-wave SSC is proportional to $D$ as $I^{s\mbox{-}\mathrm{wave}}_s \propto D$\cite{kwon}. Since the transmission coefficient $D$ is proportional to the square of the tunneling amplitude $t$, one can expect that $I^{p\mbox{-}\mathrm{wave}}_s\propto t$ and $I^{s\mbox{-}\mathrm{wave}}_s\propto t^2$. As shown in Fig. \ref{fig:hopping}, these analytical formulas can be confirmed by the numerical results obtained via the Green's function method.

For the $p$-wave spin Josephson junction, the spin Josephson current comes from the contribution of the topological edge modes carrying spin $\hbar/2$. Therefore, the spin Josephson current is proportional to the single-particle tunneling amplitude $t$, other than the exciton (two-particle) tunneling amplitude. By contrast, in the $s$-wave spin Josephson junction, only the excitons in the bulk states carrying spin $2(\hbar/2)$ contribute. As a result, the spin Josephson current here is proportional to the square of single-particle tunneling amplitude. Similarly, for a topologically trivial $p$-wave spin Josephson junction that the topological edge modes are absent, the spin Josephson current is entirely induced by the bulk states and hence it is also proportional to the square of single-particle tunneling amplitude. 



\section{$s$-WAVE SPIN JOSEPHSON JUNCTION WITH A QUANTUM DOT}\label{appendix:QD}
 
Considering the spin Josephson junction in which two $s$-wave SSCs are coupled through a spin-$1/2$ quantum dot (QD) so that the $H_t$ in Eq. (\ref{Ham_tunnel}) is replaced by 

\begin{equation}
H_t = H_{\mathrm{QD}} + \sum_{\beta=L,R} \Phi^\dagger_c \hat{t}_{\beta C} \Phi_\beta
\end{equation}

\begin{figure}[t]
\centering
\begin{minipage}[t]{6cm}
\centering
\includegraphics[height=3.75cm]{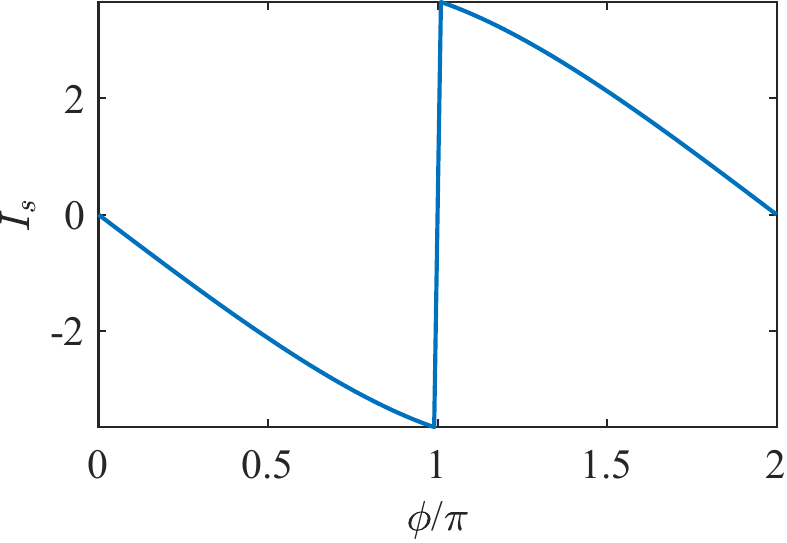}
\end{minipage}
\caption{The spin Josephson current $I_s$ with respect to the phase difference $\phi$ for a spin Josephson junction composed of two $s$-wave SSCs and a spin-$1/2$ quantum dot. The parameters chosen are $m_d = 0$, $t_0 = 0.5\Delta$, and $t_{\mathrm{LC}} = t_{\mathrm{RC}} = 0.5\Delta$.}
\label{fig:sjosdot}
\end{figure}

\noindent Here, the spin-$1/2$ QD is described by
	 
\begin{equation}
H_{\mathrm{QD}} = \Phi^\dagger_c\begin{pmatrix}
m_d&0\\
0&-m_d
\end{pmatrix} \Phi_c
\end{equation}

\noindent where $\Phi_{c} \equiv \begin{pmatrix} c_{\uparrow e}, &c^\dagger_{\uparrow h} \end{pmatrix}^\mathrm{T}$, $\hat{t}_{\beta C} \equiv t_{\beta C} \sigma_z$ is the tunneling amplitude between the QD and the left/right SSCs ($\beta = L,R$), and $m_d$ is the Zeeman splitting term for the QD. We choose this spin-$1/2$ QD as the central region and calculate the spin Josephson current adopting the Green's function method used in Sec. \ref{sec:GF}.

As shown in Fig. \ref{fig:sjosdot}, when $m_d=0$, the spin Josephson current also exhibits an abrupt jump at $\phi=\pi$
. This behavior arises because the two energy levels corresponding to the spin-up and spin-down states in the QD cross at $E=0$. The sudden jump of the spin Josephson current here at $\phi=\pi$ is accompanied with a spin flip in the QD. This phenomenon is similar to the energy level crossing of edge modes in a $p$-wave spin Josephson junction, which results in a sudden jump in the spin Josephson current, as has been discussed in detail in the main text.


\bibliography{reference_fsj}
\end{document}